# The Relationship Between Emotion Models and Artificial Intelligence

Christoph Bartneck[1], Michael Lyons[2], Martin Saerbeck[1,3]

[1] Department of Industrial Design
Eindhoven University of Technology
Den Dolech 2, 5600MB Eindhoven, The Netherlands

[2] College of Image Arts and Sciences
Ritsumeikan University
56-1 Tojiin, Kitamachi, Kita-ku, Kyoto, Japan, 603-8577

[3] Philips Research
Connected Consumer Services Group
High Tech Campus 34, 5656AE Eindhoven, The Netherlands

c.bartneck@tue.nl, michael.lyons@gmail.com, martin. saerbeck@philips.com

**Abstract.** Emotions play a central role in most forms of natural human interaction so we may expect that computational methods for the processing and expression of emotions will play a growing role in human-computer interaction. The OCC model has established itself as the standard model for emotion synthesis. A large number of studies employed the OCC model to generate emotions for their embodied characters. Many developers of such characters believe that the OCC model will be all they ever need to equip their character with emotions. This study reflects on the limitations of the OCC model specifically, and on the emotion models in general due to their dependency on artificial intelligence.

## 1 Introduction

Marvin Minsky boldly stated that "The question is not whether intelligent machines can have any emotions, but whether machines can be intelligent without any emotions" [1]. In this study, I will reflect on the relationship between emotion modeling and artificial intelligence and show that Minsky's question is still open. Emotions are an essential part of the believability of embodied characters that interact with humans [2-4]. Characters need an emotion model to synthesize emotions and express them. The emotion model should enable the character to argue about emotions the way humans do. An event that upsets humans, for example the loss of money, should also upset the character. The emotion model must be able to evaluate all situations that the character might encounter and must also provide a structure for variables influencing the intensity of an emotion. Such an emotion model enables the character to show the right emotion with the right intensity at the right time, which is necessary for the convincingness of its emotional expressions [5]. Creating such an emotion model is a daring task and in this section I will outline some of its problems. In particular, I will argue for the importance of the context in which the emotion model operates.

Emotions are particularly important for conversational embodied characters, because they are an essential part of the self-revelation feature of messages. The messages of human communication consist of four features: facts, relationship, appeal and self-revelation [6]. The inability of a conversational character to reveal its emotional state would possibly be interpreted by the user as missing sympathy. It would sound strange if the character, for example, opened the front door of the house for the user to enter and spoke with an absolute monotonous voice: "Welcome home".

## The OCC Model

From a practical point of view, the developer of a screen character of robot is wise to build upon existing models to avoid reinvent the wheel. Several emotion models are available [7, 8]. However, Ortony, Clore and Collins [9] developed a computational emotion model, that is often referred to as the OCC model, which has established itself as the standard model for emotion synthesis. A large number of studies employed the OCC model to generate emotions [2-4, 10, 11]. This model specifies 22 emotion categories based on valenced reactions to situations constructed either as being goal relevant events, as acts of an accountable agent (including itself), or as attractive or unattractive objects (see Figure 1). It also offers a structure for the variables, such as likelihood of an event or the familiarity of an object, which determines the intensity of the emotion types. It contains a sufficient level of complexity and detail to cover most situations an emotional interface character might have to deal with.

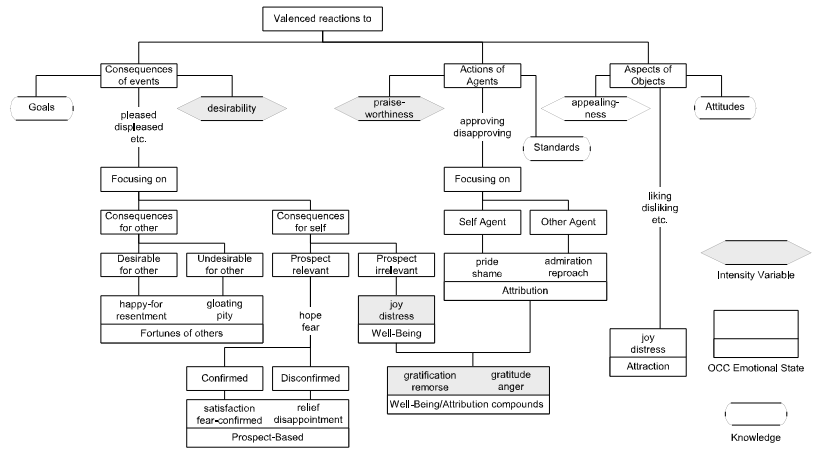

**Fig. 1.** The OCC model of emotions.

When confronted with the complexity of the OCC model many developers of characters believe that this model will be all they ever need to add emotions to their character. Only during the development process the missing features of the model and the problem of the context become apparent. These missing features and the context in which emotions arise are often underestimated and have the potential to turn the character into an unconvincing clown. I will point out what the OCC model is able to do for an embodied emotional character and what it does not.

The OCC model is complex and this paper discusses its features in terms of the process that characters follow from the initial categorization of an event to the resulting behavior of the character. The process can be split into four phases:

1. Categorization - In the categorization phase the character evaluates an event, action or object, resulting in information on what emotional categories are affected.
2. Quantification - In the quantification phase, the character calculates the intensities of the affected emotional categories.
3. Interaction - The classification and quantification define the emotional value of a certain event, action or object. This emotional value will interact with the current emotional categories of the character.
4. Mapping - The OCC model distinguishes 22 emotional categories. These need to be mapped to a possibly lower number of different emotional expressions.

**Categorization**

In the categorization phase an event, action or object is evaluated by the character, which results in information on what emotional categories are affected. This categorization requires the character to know the relation of a particular object, for example, to its attitudes. Depending on this evaluation either the "love" or "hate" emotional category will be affected by the object.

Consider this example: a character likes bananas and the user gives him a whole bunch. The character will evaluate the consequences of the event for the user, which results in pity, since the user has a whole bunch of bananas less. It will also evaluate the consequences of the event for itself, which results in satisfaction because it received a bunch of bananas. Next, it evaluates the action of the user, which results in admiration and finally the aspect of the object, which results in love. It appears that ironic that the category "love" is being used in the OCC model only for objects, since the more important usage for this word is certainly found in human-human relationships.

To do this classification the character needs an extensive amount of knowledge. First, it needs to know its relationship to the user, which was assumed to be good. Hence, pity is triggered and not resentment. Moreover, it needs to know what this event means to the user. Otherwise the character's happy-for category might be triggered (User Model). Second, it needs to have a goal "staying alive" to which the bananas contribute (Goals). Third, it needs to know what to expect from the user. Only knowing that the user does not have to hand out bananas every other minute the character will feel admiration (Standards). Last, it needs to know that it likes bananas (Attitudes).

The standards, goals and attitudes of the character that the OCC model requires need to be specified, organized and stored by the designer of the character. A new character knows even less than a newborn baby. It does not even have basic instincts. One way to store this knowledge could be an exhaustive table in which all possible events, actions and objects that the character might encounter are listed together with information on which emotional categories they affect and how their intensity may be calculated. This approach is well suited for characters that act in a limited world. However, it would be rather difficult, for example, to create such an exhaustive list for all the events, actions and objects that the character might encounter at the home of the user. With an increasing number of events, actions and objects, it becomes necessary to define abstractions. The bananas could be abstracted to food, to which also bread and coconuts belong. The categorization for the event of receiving food will be the same for all types of food. Only their intensity might be different, since a certain food could be more nutritious or tasty. However, even this approach is inherently limited. The world is highly complex and this approach can only function in very limited "cube" worlds.

This world model is not only necessary for the emotion model, but also for other components of the character. If, for example, the character uses the popular Belief, Desires and Intention (BDI) architecture [12], then the desires correspond to the goals of the emotion model. The structure of the goals is shared knowledge. So are the standards and attitudes. The complexity of the OCC model has a direct influence on the size of the required world model. However, the AI community has long given up the hope to be able to create extensive world models, such as the widely known Cyc database. The amount of information and its organization appears overwhelming. Only within the tight constraints of limited worlds was it possible so far to create operational world models.

As mentioned above, the OCC model distinguishes 22 emotional categories (see Figure 1). This rather cumbersome and to some degree arbitrary model appears to be too complex for the development of believable characters [13]. The OCC model was created to model human emotions. However, it is not necessary to model a precise human emotion system to develop a believable character. A "Black Box" approach [14] appears to be sufficient. The purpose of this approach is to produce outcomes or decisions that are similar to those resulting from humans, disregarding both the processes whereby these outcomes are attained as well as the structures involved. Such a "Black Box" approach is more suitable, particularly since the sensory, motoric and cognitive abilities of artificial characters are still far behind the ones of humans. The characters emotion system should be in balance with its abilities. Several reason speak for a simplification of the OCC model.

First, only those emotional categories of the OCC model should be used that the character can actually use. If a character uses the emotional model only to change its facial expression then its emotion categories should be limited to the ones it can express. Elliot [2] implemented all 22 emotional categories in his agents because they were able to communicate each and every one to each other. This is of course only possible for character-character interaction in a virtual world. It would be impossible for characters that interact with humans, since characters are not able to express 22 different emotional categories on their face. Ekman, Friesen and Ellsworth [15] proposed six basic emotions that can be communicated efficiently and across cultures through facial expressions.

Second, some emotional categories of the OCC model appear to be very closely related to others, such as gratitude and gratification, even thought the conditions that trigger them are different. Gratification results from a praiseworthy action the character did itself and gratitude from an action another character did. It is not clear if such a fine grained distinction has any practical advantages for the believability of characters.

Last, if the character does not have a user model then it will by definition not be able to evaluate the consequences of an event for the user. In this case, the "fortunes of others" emotional categories would need to be excluded. Ortony acknowledged that the OCC model might be too complex for the development of believable characters [13]. He proposed to use five positive categories (joy, hope, relief, pride, gratitude and love) and five negative categories (distress, fear, disappointment remorse, anger and hate). Interestingly, he excluded the emotional categories that require a user model. These ten emotional categories might still be too much for a character that only uses facial expressions. Several studies simplified the emotional model even further to allow a one-to-one mapping of the emotion model to the expressions of the character [3, 16].

**Quantification**

The intensity of an emotional category is defined separately for events, actions and objects. The intensity of the emotional categories resulting from an event is defined as the desirability and for actions and objects praiseworthiness and appealingness respectively (see Figure 1). One of the variables that is necessary to calculate desirability is the hierarchy of the character's goals. A certain goal, such as downloading a certain music album from the internet, would have several sub goals, such as download a specific song of that album. The completed goal of downloading of a whole album will evoke a higher desirability than the completed goal of downloading of a certain song, because it is positioned higher in the hierarchy. However, events might also happen outside of the character's current goal structure. The character needs to be able to evaluate such events as well. Besides the goal hierarchy, the emotion model also needs to keep a history of events, actions and objects. If the user, for example, gives the character one banana after the other in a short interval then the desirability of each of these events must decrease over time. The character needs to be less and less enthusiastic about each new banana. This history function is not described in the original OCC model, but plays an important role for the believability of the character. The history function has another important advantage. According to the OCC model, the likelihood of an event needs to be considered to calculate its desirability. The history function can help calculating this likelihood. Lets use the banana example again: The first time the character receives a banana, it will use its default likelihood to calculate the desirability of the event. When the character receives the next banana, it will look at the history and calculate how often it received a banana in the last moments. The more often it received a banana in the past the higher is the likelihood of this event and hence the lower is its desirability. After a certain period of not receiving any bananas the likelihood will fall back to its original default value. This value should not be decreased below its default value, because otherwise the character might experience an overdose of desirability the next time it receives a banana. Another benefit of the history function is the possibility to monitor the progress the character makes trying to achieve a certain goal. According to the OCC model, the effort and realization of an event needs to be considered to calculate its desirability. The history function can keep track of what the character has done and hence be the base for the calculation of effort and realization.

**Mapping**

If the emotion model has more categories than the character has abilities to express them, the emotional categories need to be mapped to the available expressions. If the character, for example, uses only facial expression then it may focus on the six basic emotions of happiness, sadness, anger, disgust, fear and surprise [15]. Interestingly, there is only one positive facial expression to which all 11 positive OCC categories need to be mapped to: the smile. Ekman [17] identified several different types of smiles but their mapping to the positive OCC categories remains unclear. The 11 negative OCC categories need to be mapped to four negative expressions: Anger, Sadness, Disgust and Fear. The facial expression of surprise cannot be linked to any OCC categories, since surprise is not considered to be an emotion in the OCC model. Even though the character might only be able to show six emotional expressions on its face, the user might very well be able to distinguish between the expression of love and pride with the help of context information. Each expression appears in a certain context that provides further information to the viewer. The user might interpret the smile of a mother next to her son receiving an academic degree as pride, but exactly the same smile towards her husband as love.

**Reflection**

The main limitation of the OCC model is its reliance on world model. Such models have only been successfully used in very limited worlds, such as pure virtual worlds in which only virtual characters operate. Furthermore, the OCC model will most likely only be one part of a larger system architecture that controls the character or robot. The emotional states of the OCC model must interact with the other states. Not only the face of the character is influenced by the emotional state of the character, but also its actions. It would be unbelievable if the character showed an angry expression on its face, but acted cooperatively. The mapping of the emotional state should be based on strong theoretical foundations. Such theoretical foundations might not be available for every action that a character might be able to execute and thus force the developer of the character to invent these mappings. This procedure has the intrinsic disadvantage that the developer might introduce an uncontrolled bias based on his or her own experiences and opinions.

Besides the actions of the character, the emotional state may also influence the attention and evaluation of events, actions and objects. In stress situations, for example, humans tend to focus their attention on the problem up to the point of "tunnel vision". [13] categorized the behavioral changes of the character through its emotional state in self-regulation (such as calming down), other-modulation (punish the other to feel better) and problem solving (try to avoid repetition). The latter will require the history function mentioned above. The emotional state of the character might even create new goals, such as calming down, which would result in actions like meditation.

## Facial Expression Synthesis

There is a long tradition within the Human-Computer Interaction (HCI) community of investigating and building screen based characters that communicate with users [18]. Recently, robots have also been introduced to communicate with the users and this area has progressed sufficiently that some review articles are available [19, 20]. The main advantage that robots have over screen based agents is that they are able to directly manipulate the world. They not only converse with users, but also perform embodied physical actions.

Nevertheless, screen based characters and robots share an overlap in motivations for and problems with communicating with users. Bartneck et al. [21] has shown, for example, that there is no significant difference in the users' perception of emotions as expressed by a robot or a screen based character. The main motivation for using facial expressions to communicate with a user is that it is, in fact, impossible not to communicate. If the face of a character or robot remains inert, it communicates indifference. To put it another way, since humans are trained to recognize and interpret facial expressions it would be wasteful to ignore this rich communication channel.

Compared to the state of the art in screen-based characters, such as Embodied Conversational Agents [18], however, the field of robot's facial expressions is underdeveloped. Much attention has been paid to robot motor skills, such as locomotion and gesturing, but relatively little work has been done on their facial expression. Two main approaches can be observed in the field of robotics and screen based characters. In one camp are researchers and engineers who work on the generation of highly realistic faces. A recent example of a highly realistic robot is the Geminoid H1, which has 13 degrees of freedom (DOF) in its face alone. The annual Miss Digital award [22] may be thought of as a benchmark for the development of this kind of realistic computer generated face. While significant progress has been made in these areas, I have not yet reached human-like detail and realism, and this is acutely true for the animation of facial expressions. Hence, many highly realistic robots and character currently struggle with the phenomena of the "Uncanny Valley" [23], with users experiencing these artificial beings to be spooky or unnerving. Even the Repliee Q1Expo is only able to convince humans of the naturalness of its expressions for at best a few seconds [24]. In summary, natural robotic expressions remain in their infancy [20].

Major obstacles to the development of realistic robots lie with the actuators and the skin. At least 25 muscles are involved in the expression in the human face. These muscles are flexible, small and can be activated very quickly. Electric motors emit noise while pneumatic actuators are difficult to control. These problems often result in robotic heads that either have a small number of actuators or a somewhat larger-than-normal head. The Geminoid H1 robot, for example, is approximately five percent larger than its human counterpart. It also remains difficult to attach skin, which is often made of latex, to the head. This results in unnatural and non-human looking wrinkles and folds in the face.

At the other end of the spectrum, there are many researchers who are developing more iconic faces. Bartneck [25] showed that a robot with only two DOF in the face can produce a considerable repertoire of emotional expressions that make the interaction with the robot more enjoyable. Many popular robots, such as Asimo, Aibo and PaPeRo have only a schematic face with few or no actuators. Some of these only feature LEDs for creating facial expressions. The recently developed iCat robot is a good example of an iconic robot that has a simple physically-animated face. The eyebrows and lips of this robot move and this allows synthesis of a wide range of expressions.

Another important issue that needs to be considered when designing the facial expression of the character is that they need to be convincing and distinct at low intensity levels. Most events that a character encounters will not trigger an ecstatic state of happiness. The evaluation of a certain event should be roughly the same as could be expected of a human and most events that humans encounter in everyday life do unfortunately not result in ecstasy. If the character managed to download a complete album of music it still did not save the world from global warming. Hence, it should only show an appropriate level of happiness.

While there is progress in the facial expressions of robot faces, we are sill facing several conceptional problems that stem from the field of Artificial Intelligence. Lets take the example of emotions that I discussed in detailed above. The emotional state of the character is defined through values for each of its emotional categories. This emotional state needs to be expressed through all available channels. A conversational embodied character, for example, needs to express its emotional state through its speech and facial expressions. It would be unconvincing if the character would smile, but speak with a monotonous voice. However, the systematic manipulation of speech to express emotions remains a challenge for the research community. Emotional facial expressions are understood better, but a fundamental questions remains. Shall the character only express the most dominant emotional category, or shall it express every category at the same time and hence show a blend of emotions. The blending of emotional expression requires a sophisticated face, such as Baldi from the CSLU Toolkit. Cartoon like characters, such as eMuu [16] or Koda's Poker Playing Agent [3] are not able to show blends and therefore they can only express the most dominant emotional category.

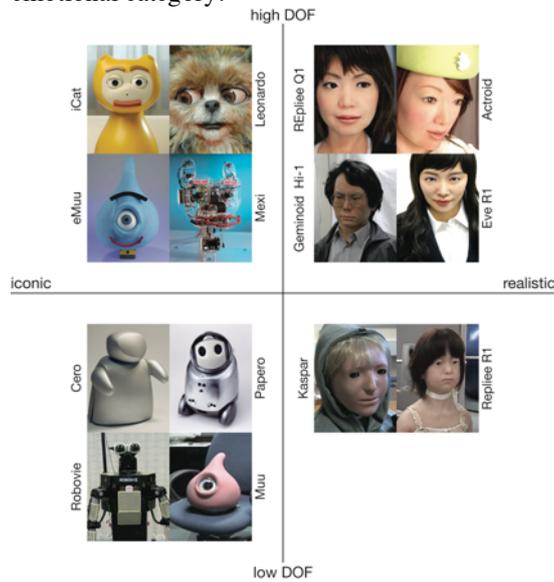

**Fig. 2.** Robots with animated faces

It becomes obvious that the problems inherited by human-robot interaction (HRI) researchers from the field of AI can be severe. Even if we neglect philosophical aspects of the AI problem and are satisfied with a computer that passes the Turing test, independently of how it achieves this, we will still encounter many practical problems. This leads us to the so-called "weak AI" position, namely claims of achieving human cognitive abilities are abandoned. Instead, this approach focuses on specific problem solving or reasoning tasks.

There has certainly been progress in weak AI, but this has not yet matured sufficiently to support artificial entities. Indeed, at present, developers of artificial entities must to resort to scripting behaviors. Clearly, the scripting approach has its limits and even the most advanced common sense database, Cyc, is largely incomplete. Emotion modeling should therefore not bet on the arrival of strong AI solutions, but focus on what weak AI solutions can offer today. Of course there is still hope that eventually also strong AI applications will become possible, but this may take a long time.

When we look at what types of HRI solutions are currently being built, we see that a large number of them do barely have any facial features at all. Qrio, Asimo and Hoap-2, for example, are only able to turn their heads with 2 degrees of freedom (DOF). Other robots, such as Aibo, are able to move their head, but have only LEDs to express their inner states in an abstract way. While these robots are intended to interact with humans, they certainly avoid facial expression synthesis. When we look at robots that have truly animated faces, we can distinguish between two dimensions: DOF and iconic/realistic appearance (see Figure 2).

Robots in the High DOF/Realistic quadrant not only have to fight with the uncannieness [26, 27] they also may raise user expectations of a strong AI which they are not able to fulfill. By contrast, the low DOF/Iconic quadrant includes robots that are extremely simple and perform well in their limited application domain. These robots lie well within the domain of the soluble. The most interesting quadrant is the High DOF/Iconic quadrant. These robots have rich facial expressions but avoid evoking associations with a strong AI through their iconic appearance. I propose that research on such robots has the greatest potential for significant advances in the use of emotions in HRI.

## Conclusion

A problem that all these artificial entities have to deal with is, that while their expression processing has reached an almost sufficient maturity, their intelligence has not. This is especially problematic, since the mere presence of an animated face raises the expectation levels of its user. An entity that is able to express emotions is also expected to recognize and understand them. The same holds true for speech. If an artificial entity talks then we also expect it to listen and understand. As we all know, no artificial entity has yet passed the Turing test or claimed the Loebner Prize. All of the examples given in Table 1 presuppose the existence of a strong AI as described by John Searle [28].

The reasons why strong AI has not yet been achieved are manifold and the topic of lengthy discussion. Briefly then, there are, from the outset, conceptual problems. John Searle [28] pointed out that digital computers alone can never truly understand reality because it only manipulates syntactical symbols that do not contain semantics. The famous 'Chinese room' example points out some conceptual constraints in the development of strong AIs. According to his line of arguments, IBM's chess playing computer "Deep Blue" does not actually understand chess. It may have beaten Kasparov, but it does so only by manipulating meaningless symbols. The creator of Deep Blue, Drew McDermott [29], replied to this criticism: "Saying Deep Blue doesn't really think about chess is like saying an airplane doesn't really fly because it doesn't flap its wings." This debate reflects different philosophical viewpoints on what it means to think and understand. For centuries philosophers have thought about such questions and perhaps the most important conclusion is that there is no conclusion at this point in time. Similarly, the possibility of developing a strong AI remains an open question. All the same, it must be admitted that some kind of progress has been made. In the past, a chess-playing machine would have been regarded as intelligent. But now it is regarded as the feat of a calculating machine – our criteria for what constitutes an intelligent machine has shifted.

In any case, suffice it to say that no sufficiently intelligent machine has yet emerged that would provide a foundation for many of the advanced application scenarios that have been imagined for emotional agents and robots. The point I hope to have made with the digression into AI is that the application dreams of researchers sometimes conceal rather unrealistic assumptions about what is possible to achieve with current technology. Emotion models heavily rely on the progress made in artificial intelligence and hence I would like to reply to Minsky's statement with a question: "Will emotional machines have intelligence?"